\documentclass{article}
\usepackage{spconf,amsmath,graphicx}
\usepackage{amsmath,graphicx}
\usepackage{amsfonts,amssymb}
\usepackage{cite}
\usepackage{booktabs}
\usepackage{subfigure}
\usepackage{verbatim}
\usepackage{multirow}

\title{Fine-tune the pretrained ATST model for sound event detection}
\name{Nian Shao$^{1, 2}$, Xian Li$^{2, 3}$, Xiaofei Li$\sthanks{Corresponding author.}^{2, 3}$}
\address{
$^{1}$ Zhejiang University, Hangzhou, China \\
$^{2}$ School of Engineering, Westlake University, Hangzhou, China \\
$^{3}$ Institute of Advanced Technology, Westlake Institute for Advanced Study, Hangzhou, China}

\begin{document}
\ninept
\maketitle
\begin{abstract}
Sound event detection (SED) often suffers from the data deficiency problem. Recent 
SED systems leverage the large pretrained self-supervised learning (SelfSL) models to mitigate such restriction, where the pretrained models help to produce more discriminative features for SED. However, the pretrained models are regarded as a frozen feature extractor in most systems, and fine-tuning of the pretrained models has been rarely studied.  
In this work, we study the fine-tuning method of the pretrained models for SED. We introduce frame-level audio teacher-student transformer model (ATST-Frame), our newly proposed SelfSL model, to the SED system. ATST-Frame was especially designed for learning frame-level representations of audio signals and obtained state-of-the-art (SOTA) performances on a series of downstream tasks. We then propose a fine-tuning method for ATST-Frame using both (in-domain) unlabelled and labelled SED data. Our experiments show that, the proposed method overcomes the overfitting problem when fine-tuning the large pretrained network, and our SED system obtains new SOTA results of 0.587/0.812 PSDS1/PSDS2 on the DCASE challenge task 4 dataset.
\end{abstract}
\section{Introduction}
\label{sec:intro}
Sound conveys a substantial amount of information about the environment. Sound event detection (SED) aims to detect sound events within an audio stream by labeling the events as well as their corresponding occurrence timestamps. Although deep neural networks have achieved significant SED performances, the high annotation budget of the task poses obstacles to the training of SED systems. Data augmentation \cite{park2019specaugment, nam2022filteraugment,zhang2017mixup}, semi-supervised learning (SemiSL) \cite{tarvainen2017mean, verma2019interpolation,koh2021sound, shao22rct},  parameter-sharing networks \cite{jiakai2018mean, zheng21skcrnn, nam22fdy} and other techniques can be used for mitigating the data deficiency problem.

The pretrained self-supervised learning (SelfSL) models are introduced to the recent SED systems. The SelfSL models learn from large-scale million-ware datasets, and are usually based on network architectures with global feature extraction capacity, such as the self-attention mechanism \cite{vaswani2017attention}. In the baseline system of DCASE2023 challenge task 4, the pretrained self-supervised BEATs model \cite{Chen2023BEATs} is integrated with a CRNN (convolutional recurrent neural network) for SED, and is regarded as an extra feature extractor. The BEATs features provide the global spectral patterns for the CRNN-based SED systems, and make the feature more discriminative for SED. With the help of BEATs features, the baseline system obtains an approximately 50\% relative improvement on the polyphonic sound detection scores (PSDSs) \cite{ebbers2022psds}, comparing with the previous CRNN baseline model. Such training paradigm becomes the fundament of top-ranked systems in the challenge \cite{Zhang2023MFD, Kim2023FDYLKA}.
However, the pretrained SelfSL models are frozen in the baseline system and most of the challenge submissions. 
Fine-tuning the pretrained SelfSL models using the (labelled) data of downstream tasks is a regular way of leveraging pretrained models \cite{li2022atst, Chen2023BEATs, li2023self}. 
This approach normally leads to improved performance compared to using the frozen pretrained models, as it allows for fine-tuning the model parameters to better suit the specific downstream tasks.
However, the fine-tuning of pretrained models in the SED systems has been rarely studied.

In this work, we incorporate the audio teacher-student transformer model (ATST) into the SED systems, named ATST-SED. We replace the BEATs model in the DCASE challenge task 4 baseline model with our newly proposed SelfSL model, frame-level ATST (ATST-Frame) \cite{li2023self}, and train/fine-tune the whole model with both labelled and (in-domain) unlabelled SED data. ATST-Frame is especially designed for learning frame-level representations of audio signals. It obtains state-of-the-art (SOTA) performances on a series of downstream tasks, including SED. Comparing with the patch-wise BEATs, ATST-Frame has a finer temporal resolution and better frame-level representations. We also propose a simple yet effective strategy to fine-tune the ATST-Frame integrated with the CRNN. Fine-tuning large model like ATST-Frame with limited amount of data would lead to the overfitting problem \cite{li2023self}. The proposed strategy applies a first-stage frozen training and a second-stage fine-tuning with heavily weighted unsupervised loss (of unlabelled data) to overcome such problem. Performance-wise, the proposed method obtains 0.587/0.812 PSDS1/PSDS2 scores on the DCASE challenge task 4 dataset, which surpasses the SOTA SED systems by a large margin, demonstrating the superiority of the proposed method.

\section{The Proposed Method}
\label{sec:method}
Polyphonic SED is a multi-class detection task to recognize the onset and offset timestamps of multiple sound events from the input audio clips. This work mainly focus on the DESED dataset \cite{turpault2019sound}. There are three types of annotated training data: the weakly labelled, strongly labelled and (in-domain) unlabelled data. To leverage all three types of the data, both supervised and unsupervised losses are used for training, optimizing the SED model in a semi-supervised manner. 
In this section, we first introduce the baseline system of DCASE2023 task 4. Next, based on the baseline system, we introduce our newly proposed ATST-Frame model to the SED system, and then present how we fine-tune the ATST-Frame model.

\begin{figure}[tb]
\begin{minipage}[b]{1.0\linewidth}
  \centering
  \centerline{\includegraphics[width=0.95\textwidth]{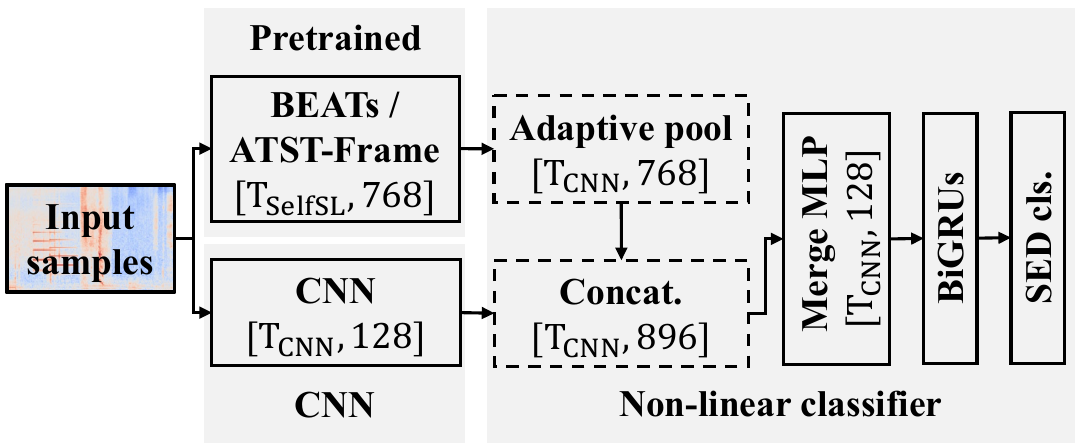}}
\end{minipage}
\caption{The architecture of the baseline and proposed SED system. The dashed blocks stand for the non-parametric modules and the solid blocks stand for the parametric modules. The feature dimensions before the merge layer are annotated, where $\text{T}_{\text{CNN}}$ and $\text{T}_{\text{SelfSL}}$ denotes the sequence length of the two modules.}
\label{fig: arch}
\vspace{-1em}
\end{figure}

\subsection{DCASE2023 baseline system and training paradigm}
The DCASE2023 challenge task 4 baseline model introduces the pretrained model, BEATs \cite{Chen2023BEATs}, to the SED systems. The architecture is shown in Fig.~\ref{fig: arch}. Comparing with the previous CRNN baseline system \cite{jiakai2018mean}, the new baseline keeps the CRNN architecture unchanged, and regards the pretrained model as an extra feature extractor running parallel to the CNN feature extractor. 

The pretrained model has a different time resolution with the CNN feature.
Specifically, the CNN feature have a time resolution of 64 ms per frame.
The pretrained model, BEATs \cite{Chen2023BEATs}, organizes the input/feature sequence patch-wisely, with a patch size of 16 frequency bands $\times$ 16 frames. Considering the input time resolution of BEATs is 10 ms per frame, then each patch feature in BEATs has a time resolution of 160 ms per patch. BEATs extracts 8 patches for each 160-ms frame, and the output sequence length of BEATs features is therefore $8 \times \frac{64}{160} = 3.2$ times longer than that of the CNN feature sequence. To align such difference, an adaptive pooling layer is applied to the BEATs feature sequence, where a window-length-varying average pooling with an adaptive stride is applied to equalize the sequence length of BEATs to the one of CNN. 


The CNN and RNN layers are trained using the DESED dataset described above in an semi-supervised learning manner, and the BEATs parameters are frozen during the training.  Binary cross entropy (BCE) loss ($\mathcal{L_\text{BCE}}$) is used for the weakly and strongly labelled data. As for the unlabelled data, MeanTeacher (MT) \cite{tarvainen2017mean} loss ($\mathcal{L_\text{MT}}$) is used. MeanTeacher holds an exponential moving average (EMA) of the training model (student model), known as the teacher model, which generates pseudo labels for the unlabelled samples. The MT loss optimizes the model by minimizing the mean-square-error between the student predictions and teacher pseudo labels. 
Mixup \cite{zhang2017mixup} is used for data augmentation with a probability of 50\%, by which two randomly selected samples as well as their corresponding labels are mixed (added) by an random interpolation factor to create new samples for the training.
The total loss ($\mathcal{L_\text{baseline}}$) can be represented as 
\begin{align}
\label{eq:loss}
\mathcal{L_\text{baseline}} = \mathcal{L}_\text{BCE} + r_\text{MT}\mathcal{L_\text{MT}},
\end{align}
where $r_\text{MT}$ stands for a ramp-up weight of the unsupervised MT loss.

The CNNs are trained using only the small task-specific DESED dataset, and mainly learn the relatively local spectral patterns. By contrast,
the SelfSL model is trained with a very large task-agnostic dataset, and equips with global feature extraction architectures such as self-attention \cite{vaswani2017attention}. The features extracted by SelfSL models and CNNs are somehow complementary to each other. As a result, when integrating the pretrained BEATs features, the baseline system obtains an over 50\% relative performance gain comparing to the CRNN system alone \cite{jiakai2018mean}. 

\begin{figure}[t]
\subfigure[BEATs features]{
\begin{minipage}[t]{0.48\linewidth}
  \centering
  \includegraphics[width=0.9\linewidth]{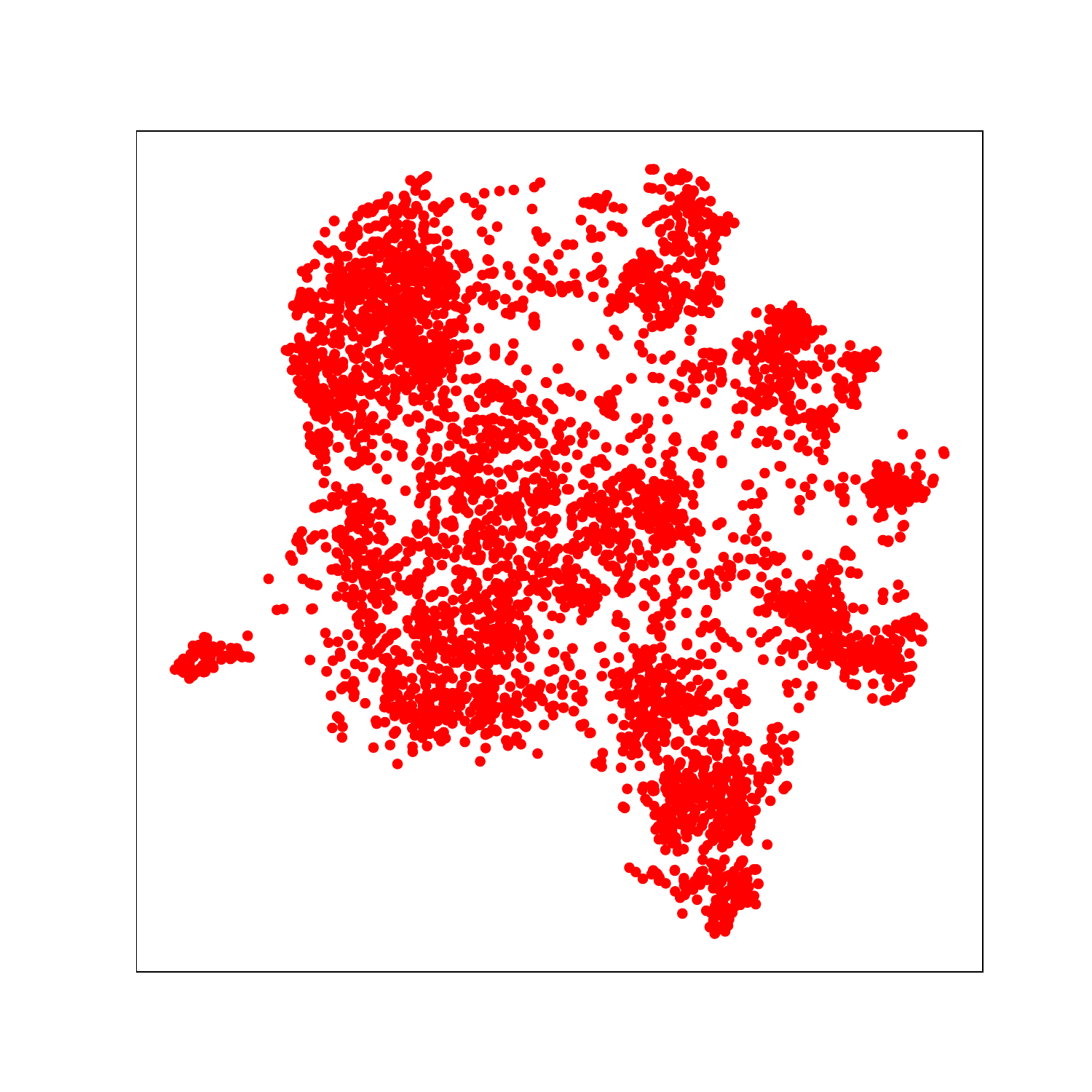}
\end{minipage}}
\subfigure[ATST-Frame features]{
\begin{minipage}[t]{0.48\linewidth}
  \centering
  \includegraphics[width=0.9\linewidth]{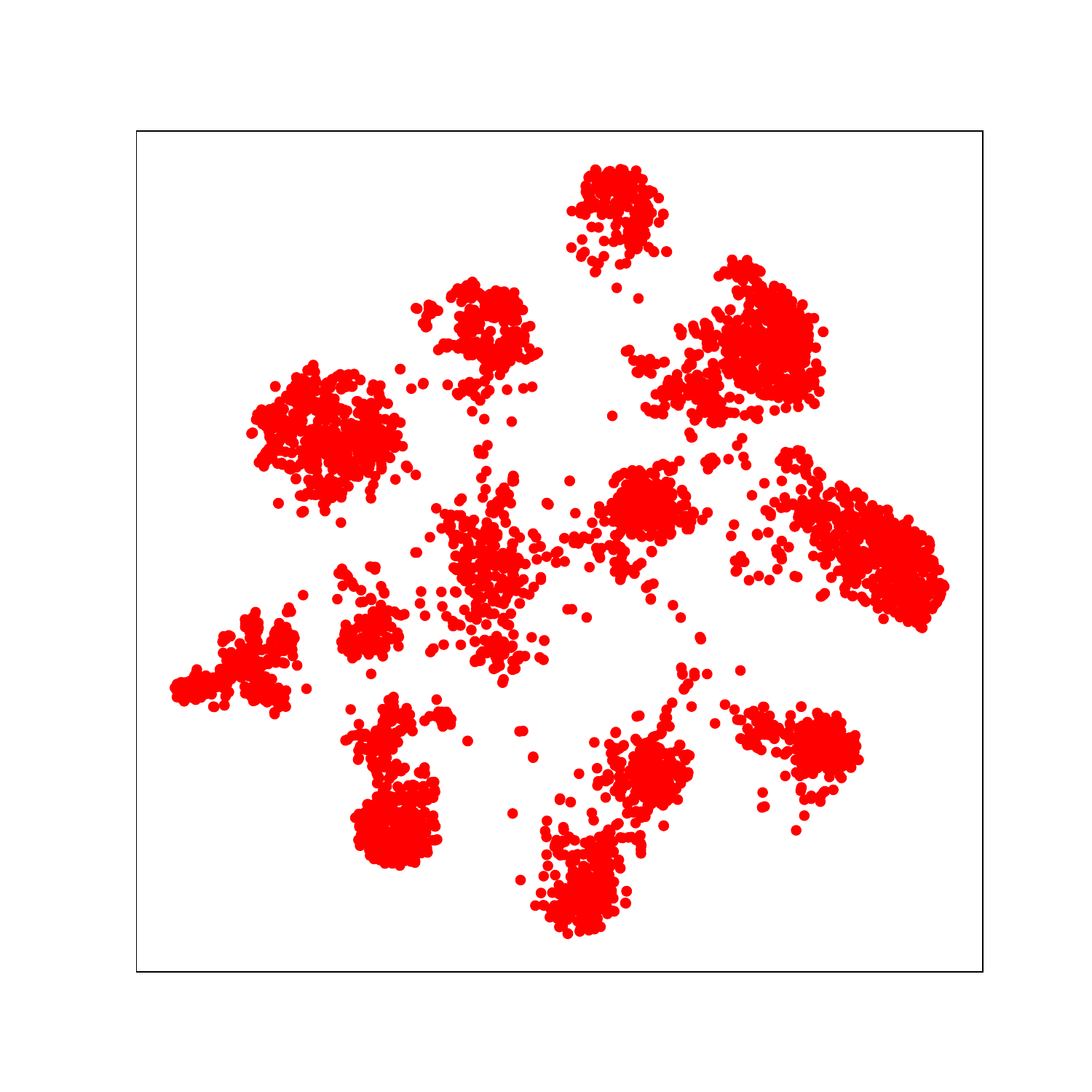}
\end{minipage}}
\caption{TSNE \cite{Hinton2008tsne} visualization on the frame-level features generated by the BEATs \cite{Chen2023BEATs} and ATST-Frame \cite{li2023self}. 
For both models, we randomly sample 1 frame-level representation from all the real audio clips in the DESED dataset.
}
\label{fig: feat_dist}
\vspace{-1em}
\end{figure}

\subsection{The proposed ATST-SED model}
In this work, we use our newly proposed ATST-Frame  model \cite{li2023self} to replace the BEATs model in the baseline system. The architecture of the overall SED system is the same as shown in  Fig.~\ref{fig: arch}. Since our system uses ATST-Frame for SED, we name it as ATST-SED. 
ATST-Frame was especially designed to produce frame-level audio representations, which is more suitable for the frame-level downstream tasks. Each ATST-Frame feature corresponds to the full spectral information of one frame. It has a finer temporal resolution (40 ms) comparing with the patch-wise BEATs features (160 ms). The ATST-Frame is even finer than the baseline SED system (64 ms). We also adopt the adaptive pooling method to align the ATST-Frame feature sequence with the CNN feature sequence. 

In \cite{li2023self}, the ATST-Frame model obtains the SOTA performance on a variety of clip-level and frame-level downstream tasks, including SED. We compare the frame-level features of ATST-Frame and BEATs, which are visualized in Fig.~\ref{fig: feat_dist} with the TSNE technique \cite{Hinton2008tsne}. For both models, we randomly sample 1 frame-level feature from all real audio clips involved in the DESED dataset. The BEATs features are obtained by averaging the 8 patch-wise embeddings of one frame. 
It can be seen that the ATST-Frame features are more densely clustered comparing with the BEATs features, illustrating an obvious linearly separable property.

\subsection{The proposed fine-tuning strategy}
The ATST-SED model can be trained in the same way as the baseline system, namely freezing the pretrained model, and only training the CRNN with the DCASE dataset. Although such way of training can obtain good SED performance, it is still sub-optimal as the pretrained model is not well adapted to the SED task. 
In this work, we propose a method to fine-tune the ATST-Frame model when it is integrated with the CRNN. 
Fine-tuning the ATST-Frame integrated in our ATST-SED system is not trivial. Regular way of fine-tuning pre-trained models uses only labelled data to fine-tune the sole pretrained model. By contrast, in our ATST-SED model, both labelled and (in-domain) unlabelled data are used in the fine-tuning, and the pretrained model is integrated with an extra CRNN. 
Our goal is to jointly train/fine-tune the CRNN and ATST-Frame networks, making the whole model optimal for SED. The proposed strategy is composed of two stages: i) freezing the ATST-Frame model, and training CRNN, as is done in the baseline system; ii) joint fine-tuning CRNN and ATST-Frame by heavily leveraging the unsupervised loss (of unlabelled data and data augmentation).  

\subsubsection{First-stage frozen training}
\begin{figure}[t]
\begin{minipage}[b]{1.0\linewidth}
  \centering
  \centerline{\includegraphics[width=0.95\textwidth]{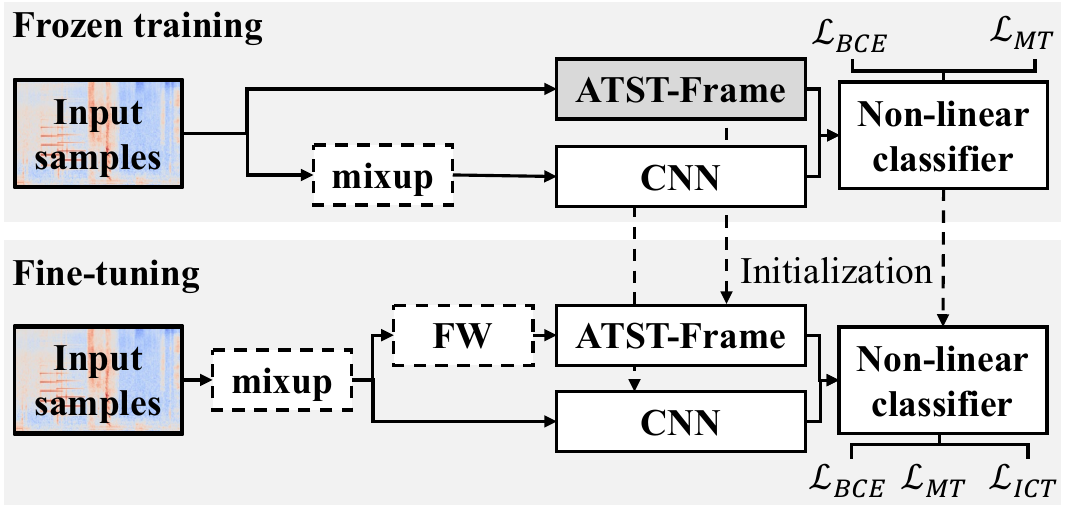}}
\end{minipage}
\caption{The flowchart of the proposed fine-tuning strategy. The gray block of ATST-Frame means it is frozen. 
}
\vspace{-1em}
\label{fig: flowchart}
\end{figure}

ATST-Frame is already a deeply pre-trained model, while the CRNN is initially not trained. It is difficult to jointly fine-tune ATST-Frame and train CRNN, as they require very different training objectives and learning rate schedule. Therefore, we first freeze the ATST-Frame and train the CRNN in the same way as the baseline system, and the CRNN learns to work in conjunction with the frozen ATST-Frame features. The detailed training strategy is illustrated in Fig.~\ref{fig: flowchart}.  After this stage, we find that it is then efficient to jointly fine-tune all the parameters of the ATST-SED model with a unified training objective and learning rate schedule. Moreover, this frozen trained model will provide reliable pseudo labels for unlabelled data, and make the next fine-tuning stage more efficient when the unsupervised losses are heavily weighted.

\subsubsection{Fine-tune with heavily weighted unsupervised loss}
\label{sec: heavy_weight_unsup}
ATST-Frame is a large model with over 80M parameters pretrained using over 2M audio clips. In the DESED dataset, the amount of labelled data is insufficient (5,048 real and 10,000 synthetic audio clips) for fine-tuning ATST-Frame. Fine-tuning the model following the baseline training paradigm would result in severe overfitting \cite{li2023self} where the model memorizes instead of learns from the data. 

In the proposed fine-tuning paradigm, we mainly rely on the unsupervised learning to overcome the overfitting problem. Specifically, we adopt data augmentation to enrich the data diversity and weight heavily on the unsupervised loss to fully leverage the larger amount of unlabelled data (14,412 real audio clips). 

As shown in Fig.~\ref{fig: flowchart}, frequency warping (FW) and mixup are applied in fine-tuning. Frequency warping was proposed in \cite{li2023self} for ATST-Frame pretraining. It randomly stretches or squeezes the spectrogram in the frequency dimension, which is efficient to improve the frequency-transformation invariance for ATST-Frame. Nonetheless, frequency warping is too difficult for the CNN module to learn, and we only apply it to the ATST-Frame inputs. 
During fine-tuning, mixup is used in conjunction with the unsupervised losses, which will be introduced later. The data augmentations are used in a random way. In each step of the fine-tuning, there is a 25\% chance of using no data augmentation, 50\% chance of using either of one augmentation and 25\% chance of using both augmentations.

Similar to the baseline training paradigm, both supervised and unsupervised learning are used in fine-tuning. The supervised BCE loss ($\mathcal{L}_\text{BCE}$) remains unchanged. For unsupervised learning, both MT ($\mathcal{L}_\text{MT}$) and interpolation consistency training (ICT) ($\mathcal{L}_\text{ICT}$) losses are used. ICT is a widely-used pseudo-labeling technique \cite{zheng21skcrnn, nam22fdy} worked in conjunction with mixup \cite{zhang2017mixup}. 
Since we use mixup with a probability of 0.5, when using mixup, the ICT loss \cite{verma2019interpolation} replaces the MT loss, and an additional weight (denoted as ${r}_{\text{ICT}}$) is applied to the ICT loss. Formally, the total loss ($\mathcal{L_\text{finetune}}$) in the proposed finetuning paradigm can be represented as: 
\begin{align}
\mathcal{L_\text{finetune}} = \mathcal{L}_\text{BCE} + r_\text{MT}\mathcal{L}_\text{MT} + r_\text{ICT}\mathcal{L}_\text{ICT}
\label{eq: finetune_loss}
\end{align}
Although the representation of $\mathcal{L}_\text{finetune}$ is similar to $\mathcal{L}_\text{baseline}$, this time the unsupervised loss weights (denoted as $r_\text{MT}, r_\text{ICT}$) are much higher. In the baseline training paradigm, the maximum value of the ramp-up weight for the MeanTeacher loss ($r_\text{MT}$ in Eq.~\ref{eq:loss}) is set to 2. In fine-tuning, the maximum values of the ramp-up weights for the MT and ICT losses ($r_\text{MT}, r_\text{ICT}$ in Eq.~\ref{eq: finetune_loss}) are set to 70 and 17.5, respectively. By such heavy weights, the gradients from the unsupervised losses dominate the training and the unsupervised data could be fully utilized. However, the unsupervised training largely relies on the quality of  pseudo labels. Thanks to the first stage frozen training, the ATST-SED model could generate reliable pseudo labels at the beginning of fine-tuning, such that ATST-Frame successfully begins to learn task-specific knowledge. During the fine-tuning, the quality of pseudo labels would keep growing. Consequently, the model performance on the SED task would keep improving until convergence.

\section{Experimental Results and Discussions}

We release our codes on our website\footnote{https://github.com/Audio-WestlakeU/ATST-SED}. All experiments are conducted on the DESED \cite{turpault2019sound} dataset, which consists of 1,578 weakly-labelled, 10,000 synthesized strongly-labelled, and 14,412 unlabelled audio clips. 3,470 extra real strongly-labelled audio clips from the AudioSet \cite{gemmeke2017audio} are used as well. We train/fine-tune both the DCASE2023 baseline model (denoted as BEATs-SED) and our proposed ATST-SED model. In pre-processing, all 10-second audio clips is resampled to 16 kHz. The input audio clips for the CNN model is frame blocked with a length of 128 ms and a hop length of 16 ms, where 128-dimensional LogMel features are extracted for each frame. As for the BEATs \cite{Chen2023BEATs} and ATST-Frame \cite{li2023self}, we directly use the pre-processing methods in their original works. 

The batch sizes for real strongly labelled, synthetic strongly labelled, real weakly labelled and real unlabelled data are set to 24, 24, 48 and 48, respectively. Adam optimizer \cite{kingma2015adam} is used. The learning rate and hyperparameter setups for both the first-stage frozen training and second-stage fine-tuning are shown in Table~\ref{tab: training_configs}. In the frozen training stage, ICT loss is not used and both pretrained models are frozen. In fine-tuning, we searched the optimal learning rates and hyperparameters for both models. ATST-Frame adopts a layer-wise learning rate decay (LLRD) setup \cite{bao2022beit, li2023self}, where the learning rate decays from $\alpha_{\text{ATST}}$ for the last Transformer block by a factor of 0.5 in each transformer block. LLRD is not applied for BEATs since it does not improve the model performance. For each parameter in the optimization, the learning rate first exponentially warms up to $\alpha$ at the $r_\text{eps}$-th epoch, and then cosine decays to $\frac{\alpha}{10}$ in the end of fine-tuning. For the weight of unsupervised losses, the unsupervised weight $r_\text{MT}$ and $r_\text{ICT}$ exponentially ramp-up from 0 to the values shown in the Table~\ref{tab: training_configs} at the $r_{\text{eps}}$-th epoch. We fine-tune the ATST-SED for 250 epochs. The system performance is evaluated through two polyphonic sound detection score (PSDS) \cite{ebbers2022psds} setups following the guidelines of DCASE2023, referred to as $\text{PSDS}_1$ and $\text{PSDS}_2$. The former metric relies more on the continuous SED results, while the latter relies more on the audio tagging accuracy. The validation metric is $\text{PSDS}_1 + \text{PSDS}_2$ of the synthetic validation set. And the model is evaluated on the DCASE2023 challenge task 4 development set.

\tabcolsep=2.5pt
\begin{table}[t]
\footnotesize
    \caption{Training configurations of ATST-SED and BEATs-SED  models in frozen and fine-tuning stages. $\alpha$ denotes the learning rate for each module. $r_\text{MT}$ and $r_\text{ICT}$ denote the maximum ramp-up weights of two losses. $r_{\text{eps}}$ is the ramp-up epochs.}
    \label{tab: training_configs}
    \centering
    \begin{tabular}{ccccccc}
    \toprule
    \textbf{Stage-Model} & $ \boldsymbol{{\alpha}_\text{CNN}}$ & $\boldsymbol{\alpha}_\text{RNN}$ & $\boldsymbol{\alpha}_{\text{SelfSL}}$ & $\boldsymbol{r}_{\text{MT}}$ & $\boldsymbol{r}_{\text{ICT}}$ & $\boldsymbol{r}_{\text{eps}}$ \\
    \midrule
    Frozen ATST-SED/BEATs-SED    & 1e-3 & 1e-3 & -    & 2  & -    & 50 \\
    Fine-tuning ATST-SED   & 2e-4 & 2e-3 & 2e-4 & 70 & 17.5 & 10 \\
    Fine-tuning BEATs-SED    & 5e-4 & 5e-4 & 5e-6 & 140 & 35 & 10 \\   
    \bottomrule
    \end{tabular}
\vspace{-1em}
\end{table}

\begin{table}[t]
    \caption{Comparison between BEATs-SED and the proposed ATST-SED in both frozen training and fine-tuning stages. \textsuperscript{$\star$} denotes the DCASE2023 baseline system. }
    \label{tab: atst_beats}
    \centering
    \begin{tabular}{clcc}
    \toprule
  \bf{Dataset} & \bf{Stage-Model} &  $\text{PSDS}_1 \uparrow$ & $\text{PSDS}_2 \uparrow$\\
  \midrule
  \multirow{2}{*}{w/o extra data} & Frozen BEATs-SED\textsuperscript{$\star$} & 0.501 & 0.755 \\
                                  & Frozen ATST-SED & 0.492 & 0.759  \\
  \midrule
  \multirow{4}{*}{with extra data} &  Frozen BEATs-SED & 0.491 & 0.781  \\
                                   &  Frozen ATST-SED  & 0.503 & 0.785  \\
                                   &  Fine-tuned BEATs-SED  & 0.521   & 0.793  \\
                                   &  Fine-tuned ATST-SED   & 0.562    & 0.802  \\
    \bottomrule
    \end{tabular}
\vspace{-1em}
\end{table}

\begin{table}[t]
    \caption{Ablation study of the proposed finetuning strategy.}
    \label{tab: ablation}
    \centering
    \begin{tabular}{lcc}
    \toprule
    \bf{Module} &  $\text{PSDS}_1 \uparrow$ & $\text{PSDS}_2 \uparrow$\\
    \midrule
    ATST-SED (proposed)                     & 0.562    & 0.802  \\
    \ \ \ \ w/o model initialization        & 0.483    & 0.791  \\
    \ \ \ \ w/o supervised losses           & 0.538    & 0.785  \\
    \ \ \ \ w/o mixup                       & 0.557    & 0.800  \\
    \ \ \ \ w/o frequency warping           & 0.556    & 0.800  \\

    \bottomrule
    \end{tabular}
\vspace{-1em}
\end{table}

\begin{table}[]
    \caption{Comparison with the SOTA SED systems. The results of other systems are quoted from their works. \textsuperscript{\dag} denotes the model trained with 7,384 extra audio clips from AudioSet \cite{gemmeke2017audio}. }
    \label{tab: comparison}
    \centering
    \begin{tabular}{lcc}
    \toprule
    \bf{Model} &  $\text{PSDS}_1 \uparrow$ & $\text{PSDS}_2 \uparrow$\\
    \midrule 
    AST-SED \cite{li2023ast}            & 0.514         & -      \\
    PaSST-SED \cite{Li2023panns}        & 0.555         & 0.791  \\
    ATST-SED (proposed)                 & \bf{0.583}    & \bf{0.810}  \\
    \midrule                                               
    FDY-LKA-BEATs \textsuperscript{\dag} \cite{Kim2023FDYLKA}  & 0.546    & 0.807  \\
    MFDConv-BEATs \textsuperscript{\dag}  \cite{Zhang2023MFD}   & 0.552    & 0.794  \\
    ATST-SED \textsuperscript{\dag} (proposed)     & \bf{0.587}    & \bf{0.812}  \\

    \bottomrule
    \end{tabular}
\vspace{-1em}
\end{table}

\subsection{Comparison between BEATs and ATST-Frame for SED}
We compare the performance of ATST-SED and BEATs-SED. Both BEATs and ATST-Frame are self-supervised trained with the balanced and unbalanced AudioSets \cite{gemmeke2017audio} and then fine-tuned on the AudioSet-2M dataset \cite{gemmeke2017audio}. 
We train the two models by using or not using the extra 3,470 real strongly-labelled audio clips, and note that the extra data is not used in the baseline system of DCASE2023 challenge task 4. The results are shown in Table~\ref{tab: atst_beats}. When the extra data is not used, BEATs-SED performs better than ATST-SED after the first-stage frozen training. However, when using extra data, the ATST-SED model shows a gain in performance in both metrics while the BEATs-SED has an unexpected performance loss on PSDS1. 
After fine-tuning, the performance measures of both models are largely promoted. This demonstrates that the proposed fine-tuning paradigm is also valid for BEATs-SED, as the optimal performance of BEATs-SED is obtained with the very large weight for unsupervised losses. The fine-tuned ATST-SED obviously outperforms the fine-tuned BEATs-SED. One major advantage of the frame-wise ATST-Frame over the patch-wise BEATs is the finer temporal resolution, and more importantly, the frame-wise sequence of ATST-Frame is automatically aligned with the SED frame sequence, while it is always difficult to align the patch-wise sequence of BEATs to the SED frame sequence, even though the model is fine-tuned with SED data. 

\subsection{Ablation study on the proposed fine-tuning paradigm}
The ablation study of the proposed fine-tuning paradigm is shown in Table~\ref{tab: ablation}. The model performance degrades when we skip the first stage frozen training, which shows it is difficult to fine-tune ATST-Frame without the help of a valid CRNN. Specifically, the quality of pseudo labels is not guaranteed when the unsupervised losses are heavily relied on. It is interesting to observe that the proposed fine-tuning method even works when the supervised loss is not used. Such experiment demonstrates the prominent effect of unsupervised learning on unlabelled data in fine-tuning. For data augmentation, both mixup and frequency warping improve the model performance. Overall, each module is beneficial to the ATST-SED performance, and is compatible with each other. Combining these modules, the proposed method significantly improves the SED performance.

\subsection{Comparison with SOTA SED systems}
Table~\ref{tab: comparison} shows the comparisons between ATST-SED and other SOTA SED systems on the DCASE2023 challenge task 4 dataset. To further improve the SED performance, the widely-used median-filter-based post-processing is also used for our ATST-SED system, and the configurations of the median filters are identical to our previous work \cite{shao22rct}.  We first compare the proposed ATST-SED with other SED systems that also based on fine-tuning large pretrained models, including AST-SED \cite{li2023ast} and PaSST-SED \cite{Li2023panns}. Comparing to these two methods, the proposed fine-tuning paradigm is simpler yet more efficient on leveraging the unlabelled data, and achieves much better SED performance. 

We also compare with the top-ranked submitted systems in DCASE2023 challenge task 4. Both of the two top-ranked systems, FDY-LKA-BEATs \cite{Kim2023FDYLKA} and MFDConv-BEATs \cite{Zhang2023MFD}, leverage extra data from the AudioSet \cite{gemmeke2017audio, hershey2021audiosetstrong}. Following the data extraction method in MFDConv-BEATs \cite{Zhang2023MFD}, 7,384 extra strongly-labelled audio clips are extracted from the strongly-labelled AudioSet and used in the fine-tuning stage of our system. Consequently, the performance of our ATST-SED is further improved, and surpasses the performance of the top-ranked systems, especially on the $\text{PSDS}_1$. Overall, by adopting our newly proposed ATST-Frame model and properly fine-tuning it, our ATST-SED system achieves new SOTA results on the DCASE2023 challenge task 4 dataset.

\section{Conclusion}
\label{sec: conclusion}
In this work, we introduce our newly proposed self-supervised learning model, ATST-Frame, to the SED system, and propose a simple yet efficient strategy to train/fine-tune the integrated ATST-Frame model. The proposed method achieves new SOTA results on the DCASE2023 challenge task 4 dataset.
Our method could potentially be applied to other pretrained models, which would form a novel training paradigm for the use of SelfSL models to SED, maybe even to other tasks.

\clearpage
\bibliographystyle{IEEE}
\bibliography{refs}

\begin{thebibliography}{10}

\bibitem{park2019specaugment}
D.~S. Park, W.~Chan, Y.~Zhang, Ch. Chiu, B.~Zoph, E.~D. Cubuk, and Q.~V. Le,
\newblock ``Specaugment: A simple data augmentation method for automatic speech
  recognition,''
\newblock in {\em Interspeech}, 2019, pp. 2613--2617.

\bibitem{nam2022filteraugment}
H.~Nam, S.-H. Kim, and Y.-H. Park,
\newblock ``Filteraugment: An acoustic environmental data augmentation
  method,''
\newblock in {\em IEEE International Conference on Acoustics, Speech and Signal
  Processing (ICASSP)}, 2022, pp. 4308--4312.

\bibitem{zhang2017mixup}
H.~Zhang, M.~Cisse, Y.~N. Dauphin, and D.~Lopez-Paz,
\newblock ``mixup: Beyond empirical risk minimization,''
\newblock in {\em International Conference on Learning Representations (ICLR)},
  2018.

\bibitem{tarvainen2017mean}
A.~Tarvainen and H.~Valpola,
\newblock ``Mean teachers are better role models: Weight-averaged consistency
  targets improve semi-supervised deep learning results,''
\newblock in {\em Advances in Neural Information Processing Systems (NIPS)},
  2017, vol.~30.

\bibitem{verma2019interpolation}
V.~Verma, A.~Lamb, J.~Kannala, Y.~Bengio, and D.~Lopez-Paz,
\newblock ``Interpolation consistency training for semi-supervised learning,''
\newblock in {\em International Joint Conference on Artificial Intelligence
  (IJCAI)}, 2019, pp. 3635--3641.

\bibitem{koh2021sound}
C.-Y. Koh, Y.-S. Chen, Y.-W. Liu, and M.~R. Bai,
\newblock ``Sound event detection by consistency training and pseudo-labeling
  with feature-pyramid convolutional recurrent neural networks,''
\newblock in {\em IEEE International Conference on Acoustics, Speech and Signal
  Processing (ICASSP)}, 2021, pp. 376--380.

\bibitem{shao22rct}
N.~Shao, E.~Loweimi, and X.~Li,
\newblock ``{RCT: Random consistency training for semi-supervised sound event
  detection},''
\newblock in {\em Interspeech}, 2022, pp. 1541--1545.

\bibitem{jiakai2018mean}
L.~JiaKai,
\newblock ``Mean teacher convolution system for dcase 2018 task 4,''
\newblock Tech. {R}ep., DCASE2018 Challenge, 2018.

\bibitem{zheng21skcrnn}
X.~Zheng, Y.~Song, I.~McLoughlin, L.~Liu, and L.-R. Dai,
\newblock ``An improved mean teacher based method for large scale weakly
  labeled semi-supervised sound event detection,''
\newblock in {\em IEEE International Conference on Acoustics, Speech and Signal
  Processing (ICASSP)}, 2021, pp. 356--360.

\bibitem{nam22fdy}
H.~Nam, S.-H. Kim, B.-Y. Ko, and Y.-H. Park,
\newblock ``{Frequency Dynamic Convolution: Frequency-Adaptive Pattern
  Recognition for Sound Event Detection},''
\newblock in {\em Interspeech}, 2022, pp. 2763--2767.

\bibitem{vaswani2017attention}
A.~Vaswani, N.~Shazeer, N.~Parmar, J.~Uszkoreit, L.~Jones, A.~N. Gomez,
  L.~Kaiser, and I.~Polosukhin,
\newblock ``Attention is all you need,''
\newblock in {\em Advances in Neural Information Processing Systems (NIPS)},
  2017, vol.~30.

\bibitem{Chen2023BEATs}
S.~Chen, Y.~Wu, C.~Wang, S.~Liu, D.~Tompkins, Z.~Chen, W.~Che, X.~Yu, and
  F.~Wei,
\newblock ``{BEAT}s: Audio pre-training with acoustic tokenizers,''
\newblock in {\em International Conference on Machine Learning (ICML)}, 2023,
  pp. 5178--5193.

\bibitem{ebbers2022psds}
J.~Ebbers, R.~Haeb-Umbach, and R.~Serizel,
\newblock ``Threshold independent evaluation of sound event detection scores,''
\newblock in {\em IEEE International Conference on Acoustics, Speech and Signal
  Processing (ICASSP)}, 2022, pp. 1021--1025.

\bibitem{Zhang2023MFD}
S.~Xiao, J.~Shen, A.~Hu, X.~Zhang, P.~Zhang, and Y.~Yan,
\newblock ``Sound event detection with weak prediction for dcase 2023 challenge
  task4a,''
\newblock Tech. {R}ep., DCASE2023 Challenge, 2023.

\bibitem{Kim2023FDYLKA}
J.-W. Kim, S.-Wo. Son, Y.~Song, H.-K Kim, I.-H. Song, and J.-E. Lim,
\newblock ``Semi-supervised learning-based sound event detection using
  frequency dynamic convolution with large kernel attention for {DCASE}
  challenge 2023 task 4,''
\newblock Tech. {R}ep., DCASE2023 Challenge, 2023.

\bibitem{li2022atst}
X.~Li and X.~Li,
\newblock ``{ATST}: Audio representation learning with teacher-student
  transformer,''
\newblock in {\em Interspeech}, 2022, pp. 4172--4176.

\bibitem{li2023self}
X.~Li, N.~Shao, and X.~Li,
\newblock ``Self-supervised audio teacher-student transformer for both
  clip-level and frame-level tasks,''
\newblock {\em arXiv preprint arXiv:2306.04186}, 2023.

\bibitem{turpault2019sound}
N.~Turpault, R.~Serizel, A.~Shah, and J.~Salamon,
\newblock ``Sound event detection in domestic environments with weakly labeled
  data and soundscape synthesis,''
\newblock in {\em Acoustic Scenes and Events 2019 Workshop (DCASE2019)}, 2019,
  p. 253.

\bibitem{Hinton2008tsne}
L.~v/d Maaten and G.~Hinton,
\newblock ``Visualizing data using t-sne,''
\newblock {\em Journal of Machine Learning Research}, pp. 2579--2605, 2008.

\bibitem{gemmeke2017audio}
J.~F. Gemmeke, D.~PW Ellis, D.~Freedman, A.~Jansen, W.~Lawrence, R.~C. Moore,
  M.~Plakal, and M.~Ritter,
\newblock ``Audio set: An ontology and human-labeled dataset for audio
  events,''
\newblock in {\em 2017 IEEE international conference on acoustics, speech and
  signal processing (ICASSP)}. IEEE, 2017, pp. 776--780.

\bibitem{kingma2015adam}
D.~P. Kingma and J.~Ba,
\newblock ``Adam: A method for stochastic optimization,''
\newblock in {\em International Conference on Learning Representations (ICLR)},
  2015.

\bibitem{bao2022beit}
H.~Bao, L.~Dong, S.~Piao, and F.~Wei,
\newblock ``{BE}i{T}: {BERT} pre-training of image transformers,''
\newblock in {\em International Conference on Learning Representations (ICLR)},
  2022.

\bibitem{li2023ast}
K.~Li, Y.~Song, L.-R. Dai, I.~McLoughlin, X.~Fang, and L.~Liu,
\newblock ``{AST-SED}: An effective sound event detection method based on audio
  spectrogram transformer,''
\newblock in {\em IEEE International Conference on Acoustics, Speech and Signal
  Processing (ICASSP)}, 2023, pp. 1--5.

\bibitem{Li2023panns}
K.~Li, P.~Cai, and Y.~Song,
\newblock ``Li {USTC} team’s submission for {DCASE} 2023 challenge task4a,''
\newblock Tech. {R}ep., DCASE2023 Challenge, June 2023.

\bibitem{hershey2021audiosetstrong}
S.~Hershey, D.~P.~W. Ellis, E.~Fonseca, A.~Jansen, C.~Liu, R.~Channing~Moore,
  and M.~Plakal,
\newblock ``The benefit of temporally-strong labels in audio event
  classification,''
\newblock in {\em IEEE International Conference on Acoustics, Speech and Signal
  Processing (ICASSP)}, 2021, pp. 366--370.

\end{thebibliography}

\end{document}